\documentclass[conference]{IEEEtran}
\IEEEoverridecommandlockouts

\usepackage{cite}
\usepackage{amsmath,amssymb,amsfonts}
\usepackage{algorithmic}
\usepackage{graphicx}
\usepackage{textcomp}
\usepackage{xcolor}
\usepackage{float}
\usepackage{booktabs}
\usepackage[hidelinks]{hyperref}

\def\BibTeX{{\rm B\kern-.05em{\sc i\kern-.025em b}\kern-.08em
    T\kern-.1667em\lower.7ex\hbox{E}\kern-.125emX}}

\begin{document}
% ADDED FOR COPYRIGHT NOTICE 
\IEEEpubid{\begin{minipage}{\textwidth}\ \\[12pt] \centering
  \copyright~2026 IEEE. Personal use of this material is permitted. Permission from IEEE must be obtained for all other uses, in any current or future media, including reprinting/republishing this material for advertising or promotional purposes, creating new collective works, for resale or redistribution to servers or lists, or reuse of any copyrighted component of this work in other works.
\end{minipage}}

\title{Financially Guided Deep Portfolio Optimization}

\author{\IEEEauthorblockN{Rahul Fernandes}
\IEEEauthorblockA{\textit{Department of Software Engineering} \\
\textit{Rochester Institute of Technology}\\
Rochester, New York, USA \\
rf4074@rit.edu}
\and
\IEEEauthorblockN{Travis Desell}
\IEEEauthorblockA{\textit{Department of Software Engineering} \\
\textit{Rochester Institute of Technology}\\
Rochester, New York, USA \\
tjdvse@rit.edu}
}
\maketitle
\begin{abstract}
Portfolio optimization in real-world financial markets is notoriously difficult due to non-stationarity, noisy data, and high transaction costs. Standard predict-then-optimize methods first forecast returns and then solve for weights, compounding prediction errors and often failing under regime shifts. We propose an end-to-end framework that directly optimizes differentiable surrogates of key financial metrics (Sharpe ratio, Omega ratio, Conditional Value-at-Risk, and risk parity), allowing neural networks to learn portfolio weights via backpropagation. Our expanding-window walk-forward procedure, applied to 50 S\&P 500 stocks from 2007 to 2023, incorporates realistic bid-ask spread costs and rebalances quarterly. On the challenging out-of-sample test period (2022--2023), the best model, an AttentionLSTM with the Omega-CVaR-RiskParity loss, achieves an annualized Sharpe of 0.29 and a total compounded return of +7.86\%, while the S\&P 500 delivers -4.52\% total compounded return and an annualized Sharpe of -0.02. This outperforms the S\&P 500 by 12.38 percentage points, while keeping tail risk (CVaR) nearly unchanged. The framework outperforms the equal-weight portfolio, S\&P 500, and traditional methods (MVP, HRP, NCO, ERC), demonstrating that embedding financial objectives directly into model training yields robust, economically meaningful outperformance even in adverse market conditions.
\end{abstract}

\section{Introduction}
Portfolio optimization is a cornerstone of finance, aiming to allocate capital across assets to maximize risk‑adjusted returns while managing constraints such as budget, leverage, and turnover. In practice, this task is fundamentally difficult because financial data is noisy, heavy‑tailed, and non‑stationary: correlations shift across regimes and estimation errors in moments can dramatically distort classical solutions. Mean‑Variance Optimization (MVO), though foundational, is especially brittle in high‑dimensional, time‑varying markets -- small errors in expected returns or covariances frequently produce extreme, concentrated allocations that perform poorly out of sample. Methods such as Hierarchical Risk Parity (HRP) and Nested Clustered Optimization (NCO) reduce some instability but remain heuristic and clustering‑based, assuming stationarity and failing to capture nonlinear, regime‑dependent dynamics. These scientific challenges make portfolio construction a promising area for methods that combine expressive function approximators with finance‑aware inductive biases.

Motivated by these challenges, we design and implement "Financially Guided Deep Portfolio Optimization", an end-to-end portfolio optimization framework that embeds financial objectives and portfolio construction into hyperparameter tuning and training of neural networks. Rather than following a predict-then-optimize pipeline (forecast returns, then solve for portfolio allocation), our models output normalized portfolio allocation weights. Our key contribution is an integrated, end‑to‑end framework that combines: (i) differentiable surrogates for financial metrics (Sharpe, Omega, CVaR, risk parity); (ii) an expanding‑window walk‑forward evaluation procedure that mimics real‑world rebalancing; and (iii) a maximin hyperparameter optimization strategy. Together, these elements form a robust pipeline for training neural networks directly on portfolio‑level objectives.

We evaluate seven neural architectures with two custom loss functions on a CRSP dataset of 50 S\&P 500 constituents spanning 2007 to 2023. An expanding-window walk-forward validation and a hyperparameter search select the most promising model-loss combinations, which are then tested on the out-of-sample 2022--2023 period. Our results show that the AttentionLSTM with the Omega-CVaR-RiskParity loss (Section \ref{subsec:losses}) achieves an annualized Sharpe ratio of +0.29, while the S\&P 500 delivers \textminus0.02. The total compounded return (net of transaction costs) is +7.86\%, beating the S\&P 500's \textminus4.52\% by 12.38 percentage points. At the same time, the model's CVaR is only \textminus2.86\%, nearly identical to the S\&P 500's \textminus2.84\%, demonstrating that the substantially higher return comes with a negligible increase in tail risk. These findings show that a relatively simple recurrent architecture, when paired with a robust end‑to‑end pipeline, can decisively outperform both traditional methods and market benchmarks.

\section{Related Work}
\IEEEpubidadjcol % ADDED FOR COPYRIGHT NOTICE
Portfolio construction research spans interconnected strands, including classical mean-variance theory and robust covariance estimation, predict-then-optimize pipelines, end-to-end neural methods, differentiable performance objectives, and evolving architectures. The following subsections summarize representative works, identify key gaps (especially in embedding diversification and risk-aware constraints into training), and show how these motivate our Financially Guided Deep Portfolio Optimization. This review focuses on robustness to estimation errors and regime shifts, contrasts heuristic/clustering-based methods with learning-based approaches, and thereby shows how our design addresses the shortcomings of prior work.

\subsection{Classical Portfolio Theory and Robust Covariance Methods}
Modern portfolio theory formalized portfolio construction as a mean-variance optimization problem \cite{markowitz1952modern}. MVO's sensitivity to estimation error and non-stationarity motivated alternative, more robust constructions. Hierarchical Risk Parity (HRP) \cite{lopez2016building} uses hierarchical clustering and quasi-diagonalization of the covariance matrix to produce more stable allocations. Related hierarchical clustering and nested clustering approaches like Nested Clustered Optimization (NCO)\cite{lopez2016robust} also seek robust, less concentrated allocations via cluster-driven allocation rules. These methods improve empirical out-of-sample stability, but remain heuristic and clustering-based, assume stationarity, and can fail to capture nonlinear temporal dependencies. We employ these methods as benchmarks in our experiments.

\subsection{Predict-then-Optimize vs. End-to-End Learning}
Machine learning solutions for portfolio optimization fall into two broad paradigms. The first, predict-then-optimize: fits forecasts (returns, risk forecasts) and then solves a downstream optimization problem; while modular, this pipeline compounds forecasting errors and often ignores allocation objectives during training \cite{ban2018machine,behera2025approach}. The second paradigm trains models end-to-end to output allocations directly, optimizing task-aligned objectives (e.g., Sharpe-style surrogates). End-to-end methods have been shown to improve backtest metrics in many settings and avoid covariance inversion issues, e.g. DELAFO\cite{cao2020delafo} and similar neural end-to-end frameworks. However, most end‑to‑end approaches concentrate on return or risk‑adjusted metrics (e.g., Sharpe) and do not explicitly enforce portfolio diversification, a gap we address with our regularization.

\subsection{Differentiable Objectives and Regularization}
A growing literature constructs differentiable surrogates for, or inspired by, investor-relevant metrics (Returns, Sharpe, Sortino, CVaR) so networks can be trained directly on portfolio-level objectives \cite{izzo2023higher, kubo2025portfolio, guo2025pursuing, HUANG2026129672}. Performance-based regularization (PBR)\cite{ban2018machine} penalizes portfolios with high estimation variability (e.g., mean/CVaR variability) to reduce estimation risk and improve empirical frontiers \cite{karoui2011performance}. Practical challenges remain; smooth surrogates can be noisy and require careful normalization, batching, and regularization, but these approaches motivate our design of differentiable loss functions that combine financial objectives such as Sharpe, Omega, CVaR, and risk parity.

\subsection{Model Architectures and Neural Architecture Search}
For temporal financial data, recurrent models (RNNs, LSTMs, GRUs), attention mechanisms, and transformer variants have been explored extensively. Hybrid architectures aim to capture local and long-range patterns; transformer-style models and temporal fusion architectures have also been adapted for multi-horizon forecasting. In particular, the Temporal Fusion Transformer (TFT) \cite{LIM20211748} uses an LSTM recurrent encoder, variable selection networks, and multi‑head attention. More recently, DeformTime \cite{shu2025deformtime} introduces a deformable attention mechanism that adaptively aligns time steps, improving the handling of misaligned or varying‑frequency data. Similarly, PatchTST \cite{nie2023patchtst} segments input into patches and applies a Transformer for efficient long‑range modeling. Both DeformTime and PatchTST have shown strong performance in time series forecasting; in this work we adapt them as benchmarks, and we incorporate the Variable Selection Network from TFT into our VSN-LSTM architecture. Neuroevolution methods like EXAMM \cite{lyu2024neuroevolution} have also been applied to evolve recurrent topologies, but architecture improvements only translate to robust allocations when paired with task‑aligned, risk‑aware objectives and realistic validation.

\section{Methodology}

\subsection{Problem Statement and Preliminaries}
Portfolio optimization is a mathematical framework used to determine the optimal allocation of capital across a set of financial assets. Its primary objective is to maximize the expected return for a specific level of risk, or conversely, to minimize risk for a target level of return.

Consider a collection of $N$ assets. Let $R_t \in \mathbb{R}^N$ denote the vector of daily returns of all assets at time $t$. At the rebalancing date, a neural network produces a vector of portfolio allocation weights $w \in \mathbb{R}^N$ satisfying
\begin{equation}
    w_{i} \ge 0 \qquad \text{and} \qquad \sum_{i=1}^{N} w_{i} = 1,
\label{eq:weights}
\end{equation}
i.e., a long-only, fully invested portfolio. The portfolio's daily return at time $t$ is given by
\begin{equation}
    r_t = w^\top R_t.
\label{eq:daily_rets}
\end{equation}

Given a historical input window of length $T_{in}$ (e.g., 180 trading days) and a future holding period of $T_{out}$ (e.g., 60 trading days), the goal is to learn a function $f_\theta$ (parameterized by $\theta$) that maps the past features to weights that maximize the risk-adjusted performance metric over the holding period $T_{out}$. In our research, the typical metrics are the Sharpe ratio and Omega ratio.

Financial returns are notoriously non-stationary and noisy, hence the direct optimization of these can lead to overfitting to a particular signal in the data, which can lead to underperforming models on out-of-sample data. We therefore augment the objective with differentiable regularizers that encourage tail-risk control (Conditional Value-at-Risk) and diversification (risk parity). The entire system is trained end-to-end via backpropagation. The key financial metrics used in this work are defined as follows:

\begin{itemize}
    \item Sharpe Ratio: Measures how much excess return (over a risk-free rate) an investment earns per unit of total risk (volatility). A higher Sharpe ratio indicates better risk‑adjusted performance.
    \item Omega Ratio: Captures the asymmetry of returns by considering the entire distribution, not just average and variance. It is defined as the ratio of the probability‑weighted gains above a given threshold to the probability‑weighted losses below that threshold. An Omega ratio $>1$ indicates that gains outweigh losses.
    \item Conditional Value-at-Risk (CVaR): Also known as Expected Shortfall, estimates the average loss that an investment may suffer on the worst days (typically the worst 5\% of trading days).
    \item Risk Parity: An asset allocation strategy that aims to distribute the overall portfolio risk evenly among its components. Instead of allocating capital based on expected returns, risk parity sets weights so that each asset contributes the same amount to the portfolio's total volatility.
    \item Information Ratio: Measures excess return relative to a benchmark per unit of active risk (also called tracking error). A higher Information Ratio indicates more consistent outperformance of the benchmark.
\end{itemize}

\subsection{Candidate Model Architectures}
We consider five neural architectures that differ in temporal modeling. All share a final softmax layer to output portfolio weights $w\in\mathbb{R}^N$.

\subsubsection{BaseLSTM} A multi‑layer LSTM whose last hidden state is passed through a linear layer to produce portfolio weights.

\subsubsection{AttentionLSTM} Extends BaseLSTM with a multi‑head self‑attention over the LSTM hidden states, followed by residual connection and layer norm. The attention output is averaged over time before the final linear layer. This allows the model to focus on the most informative time steps, potentially improving robustness.

\subsubsection{InvertedAttentionLSTM} This reverses the conventional role of time and feature dimensions. After an LSTM, the hidden tensor is transposed to swap time and feature dimensions. Multi‑head attention is applied across features, treating each hidden unit as a token. The result is mean‑pooled and linearly projected to portfolio weights.

\subsubsection{TemporalTransformer} Combines an LSTM for local smoothing with a Transformer encoder for global dependencies. A learnable positional embedding is added to the LSTM output, then passed through several Transformer layers. The time‑averaged output feeds the final linear layer. This hybrid design leverages the LSTM's ability to filter high‑frequency noise while the Transformer attends to global patterns over the entire window.

\subsubsection{VSN-LSTM} Inspired by the Temporal Fusion Transformer \cite{LIM20211748}, a Variable Selection Network learns per‑feature importance via a small MLP with sigmoid gating. Weighted features are processed by an LSTM, then optionally by temporal attention, before the final linear layer.

\subsection{Loss Function Construction}
\label{subsec:losses}
At the core of our method are differentiable loss functions that directly optimize portfolio performance metrics, rather than predicting returns. By embedding portfolio construction objectives directly into the loss, we train the models end‑to‑end to produce portfolio weights that maximize a chosen risk‑adjusted performance measure, while simultaneously regularizing tail risk and enforcing structural diversification. All loss functions are computed on the portfolio's daily returns ($r = [r_1,\dots, r_{T_{out}}]$, derived from the weights and asset returns) over the 60-day holding period $T_{out}$. They are composed of three additive terms:

\begin{itemize}
    \item \textbf{Primary Objective}: the main performance metric to be maximized (higher is better).
    \item \textbf{Tail‑Risk Regularizer}: penalizes large losses, encouraging robustness to extreme events.
    \item \textbf{Diversification Regularizer}: enforces that risk is evenly distributed across assets, promoting diversification.
\end{itemize}

\begin{figure}[htbp]
\centering
\includegraphics[width=\columnwidth, height=1.4in]{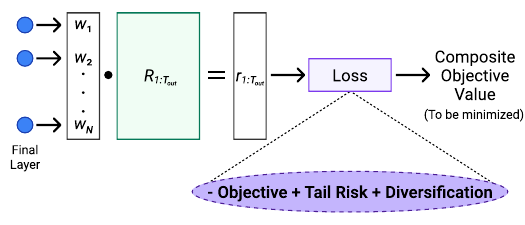}
\caption{Typical Forward Pass}
\label{fig:forward-pass}
\end{figure}

\subsubsection{\textbf{Smooth Sharpe Objective}} The Sharpe ratio can be mathematically defined as:
\begin{equation}
    S = \frac{\mu_p - R_f}{\sigma_p},
\label{eq:typ_sharpe}
\end{equation}
where $\mu_p$ is the mean portfolio return, $R_f$ is the risk‑free rate, and $\sigma_p$ is the standard deviation of portfolio returns (volatility).
% \begin{equation}
%     \mu_p = \frac{1}{T} \sum_{t=1}^T r_t, \quad \sigma_p = \sqrt{\frac{1}{T}\sum_{t=1}^T\left(r_t-\mu_p\right)^2}
% \label{eq:sharpe_mean_std}
% \end{equation}
To make it smooth, differentiable and always positive, we apply a softplus transformation and take the negative log. Since our dataset consists of only stocks and no other form of assets, which could be risk free, the effective risk free rate for the calculation of the Sharpe ratio is zero.
Hence, the smooth Sharpe loss objective term is,
\begin{equation}
    \mathcal{L}_{\text{Sharpe}} = -\log\left[ \operatorname{softplus} \left(\frac{\mu_p}{\sigma_p + \varepsilon} \right) \right],
\label{eq:sharpe_obj}
\end{equation}
where $\operatorname{softplus}(x)=\frac{1}{\beta}\log(1+e^{\beta x})\, \text{ with } \beta=1$.

\subsubsection{\textbf{Smooth Omega Objective}} The Omega ratio can be mathematically defined as:
\begin{equation}
    \Omega = \frac{\frac{1}{T_{out}}\sum_{t=1}^{T_{out}} \max(0, r_t - \theta)}{\frac{1}{T_{out}}\sum_{t=1}^{T_{out}} \max(0, \theta - r_t) + \varepsilon},
\label{eq:typ_omega}
\end{equation}
where $\theta$ is the threshold (here $\theta = 0$).
To obtain a smooth differentiable version, we replace the hard max with a softplus function and take the negative log:

\begin{equation}
    \mathcal{L}_{\text{Omega}} = -\log\left[\frac{\frac{1}{T_{out}}\sum_{t=1}^{T_{out}} \operatorname{softplus}(r_t - \theta)}{\frac{1}{T_{out}}\sum_{t=1}^{T_{out}} \operatorname{softplus}(\theta - r_t) + \varepsilon}\right].
\label{eq:omega_obj}
\end{equation}

\subsubsection{\textbf{Differentiable CVaR Regularizer}} The CVaR  can be mathematically defined as:
\begin{equation}
    \operatorname{CVaR}_{\alpha}(L) = \mathbb{E}\bigl[L \;\big|\; L \geq \operatorname{VaR}_{\alpha}(L)\bigr],
\label{eq:typ_cvar}
\end{equation}
where, $L = -r$ is the portfolio loss (negative return) and $\operatorname{VaR}_{\alpha}(L)$ is the value‑at‑risk at confidence level $\alpha$. The Rockafellar‑Uryasev \cite{rockafellar2000optimization}\cite{rockafellar2002conditional} formula gives a differentiable approximation:
\begin{equation}
    \operatorname{CVaR}_{\alpha}(r) = \zeta + \frac{1}{\alpha}\mathbb{E}\left[\max(0, -r-\zeta)\right],
\label{eq:ru_cvar}
\end{equation}
where $\zeta = \operatorname{VaR}_{\alpha}(-r)$ is the value‑at‑risk of the losses. We use a smooth softplus surrogate for the $\max$ and estimate $\zeta$ without gradients (stop‑gradient). Hence, the final term is,
\begin{equation}
    \mathcal{L}_{\text{CVaR}} = \zeta + \frac{1}{\alpha} \frac{1}{T_{out}} \sum_{t=1}^{T_{out}} \operatorname{softplus}(-r_t - \zeta),
\end{equation}
where $\zeta = \text{Quantile}_{1-\alpha}(-r)$ (the value‑at‑risk of the losses). We then normalize this by the portfolio's standard deviation ($\sigma_p$) to make it scale invariant. This regularizer penalizes large negative returns and encourages the model to avoid tail risk.

\subsubsection{\textbf{Risk Parity Regularizer}}
Let $\Sigma$ be the covariance matrix of the $N$ assets over the holding period $T_{out}$, and let $w$ be the portfolio weights vector. The portfolio variance is $\sigma_p^2 = w^{\top}\Sigma w$. The risk contribution of each asset $i$ is $RC_i = w_i(\Sigma w)_i$, where $(\Sigma w)_i = \sum_{j=1}^{N} \sigma_{ij} w_j$. This satisfies $\sum_{i=1}^{N} RC_i = \sigma_p^2$. Risk parity requires $RC_i = \sigma_p^2 / N$ for all $i$. Equivalently,
\begin{equation}
    w_i (\Sigma w)_i = \frac{w^{\top}\Sigma w}{N} \quad \forall i.
\label{eq:typ_rp}
\end{equation}
For the regularizer, we minimize the squared deviation:
\begin{equation}
    \mathcal{L}_{\mathrm{RP}} = \sum_{i=1}^{N} \left( RC_i - \frac{w^{\top}\Sigma w}{N} \right)^2.
\label{eq:rp_reg}
\end{equation}
To improve numerical stability, we use a shrunk covariance matrix (linear shrinkage towards a diagonal matrix) and make the loss scale‑invariant by dividing by $(w^\top\Sigma w)^2$.

\subsubsection{\textbf{Custom Loss Functions}}
\begin{itemize}
    \item \textbf{CustomLossA} (Sharpe-based):
    \begin{equation}
        \mathcal{L}_{\text{CustomLossA}} = \mathcal{L}_{\text{Sharpe}} + \lambda_{\text{CVaR}} \cdot \mathcal{L}_{\text{CVaR}} + \lambda_{\text{RP}} \cdot \mathcal{L}_{\text{RP}}
    \label{eq:loss_A}
    \end{equation}
    \item \textbf{CustomLossB} (Omega-based): 
    \begin{equation}
        \mathcal{L}_{\text{CustomLossB}} = \mathcal{L}_{\text{Omega}} + \lambda_{\text{CVaR}} \cdot \mathcal{L}_{\text{CVaR}} + \lambda_{\text{RP}} \cdot \mathcal{L}_{\text{RP}}
    \label{eq:loss_B}
    \end{equation}
\end{itemize}

Both losses are minimized during training. The hyperparameters control the trade‑off between maximizing risk‑adjusted return, controlling tail risk, and enforcing diversification. We refer to these as CustomLossA (Sharpe-CVaR-RiskParity) and CustomLossB (Omega‑CVaR‑RiskParity) respectively. We initially explored 16 combinations of financial metrics (log Returns, Sharpe, Omega, Sortino, Calmar, CVaR, MDD, Risk Parity, HHI, Entropy) as regularizers or primary objectives. Most led to overfitting or poor validation performance. The two loss functions that consistently performed well, CustomLossA and CustomLossB, are used in the experiments.

\subsection{Expanding-Window Walk-Forward Procedure} 
\label{subsec:wfv}
To simulate realistic portfolio management, we adopt a buy-and-hold strategy for 60 trading days ($\approx1$ quarter), rebalancing every quarter. We evaluate all models using an expanding-window walk-forward procedure to simulate this strategy. The data set is divided chronologically into an initial training period, a validation period, and a test period. The validation and test periods contain eight steps ($K=8$). The expanding-window walk-forward is executed as follows. \\
At each walk-forward step $k=1,2,\dots,K$:
\begin{enumerate}
    \item \textbf{Expansion:} The training set is expanded by adding the data containing all features (not just returns) for the previous 60 day holding period, if $k>1$ (No expansion for the first step).
    \item \textbf{Training:} We generate training samples using rolling windows from the cumulative training set (or initial training set if $k=1$), i.e., all available historical data up to the start of the current window. Given an input window length of $T_{in}=180$ days and an output (holding) window of $T_{out}=60$ days, we slide a window with stride $=$ 1 over the available data. This produces a large number of overlapping input--output pairs: each input is a $180 \times F$ matrix of features (where $F=251$), and each output is the corresponding $60 \times N$ matrix of asset returns (where $N=50$). The model is trained on this augmented and cumulative training data.
    \item \textbf{Prediction:} Using the most recent 180 days ($T_{in}$) as the inference input window, the model outputs a vector of portfolio weights $w^{(k)} \in \mathbb{R}^N$ for the next 60 days ($T_{out}$). These weights are to be held constant for the entire output window.
\end{enumerate}
\begin{figure}[htbp]
\centering
\includegraphics[width=\columnwidth]{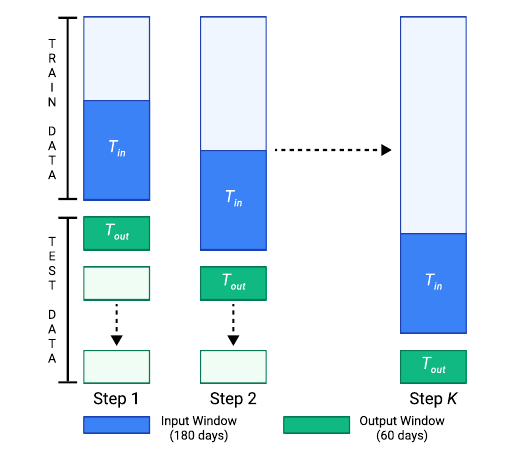}
\caption{Walk-Forward Procedure}
\label{fig:wfv}
\end{figure}

This process repeats for $K$ steps, covering the entire validation or test period. The output windows are non‑overlapping, so each step provides an independent out‑of‑sample performance estimate. For each step $k$, the portfolio weights produce a daily returns series $r^{(k)} = R^{(k)} w^{(k)}$, where $R^{(k)}$ denotes the asset returns for $T_{out}$. Transaction costs are incorporated into the evaluation: at each rebalancing, we compute the total cost using the bid‑ask spread (BA spread) of each asset on the first day of the window, and subtract it from the first day's portfolio return. The exact cost model is detailed in Section \ref{subsec:transaction_costs}. All reported portfolio performance metrics are computed on the net-of-cost daily returns. The same walk-forward procedure is used during hyperparameter tuning and during the final testing.

% \begin{figure}[htbp]
% \centering
% \includegraphics[width=\columnwidth]{images/wfv.pdf}
% \caption{Walk-Forward Procedure}
% \label{fig:wfv}
% \end{figure}

\subsection{Hyperparameter Optimization}
All model hyperparameters (learning rate, weight decay, hidden size, number of layers, attention heads, dropout, loss regularization weights, epochs, etc.) are tuned using Optuna \cite{ozaki2025optunahub}, a Bayesian optimization framework. For each hyperparameter configuration, we run a full expanding‑window walk‑forward on the validation set. The tuning objective value for a trial is computed as follows:

\subsubsection{\textbf{Per-Step Excess Returns and Information Ratio}} At each walk-forward step $k=1,\dots,K$, the model's predicted portfolio weights yield a 60-day portfolio returns series $r^{(k)}$. We subtract the daily returns of the equal-weight benchmark $b^{(k)}$ for the same 60 days to obtain the excess returns:
\begin{equation}
    \alpha^{(k)}_t = r^{(k)}_t - b^{(k)}_t, \quad t = 1,\dots, 60.
\label{eq:exc_rets}
\end{equation}
The Information Ratio ($ir$) for step $k$ is:
\begin{equation}
    ir^{(k)} = \frac{\bar{\alpha}^{(k)}}{\sigma_{\alpha}^{(k)}},
\label{eq:info_ratio}
\end{equation}
where $\bar{\alpha}^{(k)}$ and $\sigma_{\alpha}^{(k)}$ are the mean and standard deviation of $\alpha^{(k)}$. This measures how consistently the model outperforms the benchmark.

\subsubsection{\textbf{Tuning Objective}} To select the hyperparameters that perform well on average while being stable across the walk-forward steps, we adopt a maximin optimization strategy, choosing configurations that maximize the 95\% lower confidence bound of the mean Information Ratio:
\begin{equation}
    \mathrm{Objective} = \bar{IR} - t_{0.95,\,K-1} \cdot \frac{\sigma_{IR}}{\sqrt{K}},
\label{eq:tune_obj}
\end{equation}
where $\bar{IR} = \frac{1}{K} \sum_{k=1}^{K} ir^{(k)}$ and $\sigma_{IR} = \sqrt{\frac{1}{K-1}\sum_{k=1}^K (ir^{(k)} - \bar{IR})^2}$ are the sample mean and standard deviation of the Information Ratios from each walk-forward step, and $t_{0.95, K-1}$ is the critical value of the Student's $t$-distribution with $K-1$ degrees of freedom for a one sided 95\% confidence interval.

\subsection{Candidate Model Selection}
Since the risk-adjusted metrics (Sharpe, Sortino, Calmar, Omega) were highly correlated, and tail metrics (CVaR, MDD) were highly correlated, we selected Sharpe ratio as the primary measure of risk-adjusted return and CVaR as the primary measure of tail risk. We then applied Pareto dominance on the validation set performance using these two metrics. This yielded a pareto front of non-dominated model-loss combinations. As a secondary check, we performed PCA on all seven metrics. PC1 captured risk-adjusted metrics and PC2 captured tail risk. By evaluating the front using PC1 and an inverted PC2, we sampled the opposite boundary of the risk--return spectrum to include a more aggressive tradeoff profile for comparison. This secondary front added two more models. The union of both fronts gave six candidates out of fourteen model-loss combinations: AttentionLSTM-CustomLossB, DeformTime-CustomLossB, TemporalTransformer-CustomLossB, PatchTST-CustomLossB, VSN-LSTM-CustomLossB, and InvertedAttentionLSTM-CustomLossA. These six were carried forward to the final test set evaluation.
% Since, the risk‑adjusted metrics (Sharpe, Sortino, Calmar, Omega) were highly correlated, and tail metrics (CVaR, MDD) were highly correlated, we selected Sharpe ratio as the primary measure of risk‑adjusted return and CVaR as the primary measure of tail risk. We then applied Pareto dominance on the validation set performance using these two metrics. This yielded a frontier of non-dominated model-loss combinations. As a secondary check, we performed PCA on all seven metrics. PC1 captured risk‑adjusted metrics and PC2 captured tail risk metrics; we used PC1 and inverted PC2 to obtain a 'higher‑is‑better' space. The Pareto frontier in this space added one more model. The union of both frontiers gave six candidates out of fourteen model-loss combinations: AttentionLSTM-CustomLossB, DeformTime-CustomLossB, TemporalTransformer-CustomLossB, PatchTST-CustomLossB, VSN-LSTM-CustomLossB, and InvertedAttentionLSTM-CustomLossA. These six were carried forward to the final test set evaluation.

\section{Implementation}
\subsection{Data and Preprocessing}
\subsubsection{Raw Data} We use daily data from the Center for Research in Security Prices (CRSP) for 50 constituent stocks of the S\&P 500 index, spanning a 16 year period. The feature set consists of five features for each stock: daily returns, change in volume, bid-ask spread, turnover and illiquidity. Also, there is an additional column providing the daily returns of the S\&P 500 index. The full timeline is split chronologically into three non‑overlapping sets:
\begin{itemize}
    \item Training set: 7 December 2007 -- 6 February 2020
    \item Validation set: 7 February 2020 -- 31 December 2021
    \item Test set: 3 January 2022 -- 29 November 2023
\end{itemize}
\subsubsection{Preprocessing} The preprocessing is as follows:

\paragraph{Extraction} From the raw data, we extract: (\textit{i}) the daily returns of all 50 stocks, (\textit{ii}) the BA spread values for each stock (needed for transaction cost calculations on the validation and test sets), and (\textit{iii}) the S\&P 500 returns column (used as a benchmark). These extracted columns are not normalized and serve as target outputs (asset returns) and auxiliary data.

\paragraph{Normalization} The original full feature set (including S\&P 500 returns) is scaled using a RobustScaler (median and interquartile range). Robust scaling is chosen because the financial data exhibited fat-tailed distributions (and skewed for turnover and BA spread). The scaler is fitted only on the training set and the same statistics are then applied to the validation and test sets. This fixed normalization avoids look‑ahead bias and was found to generalize better than re‑fitting the scaler at each walk‑forward step.

\subsection{Neural Network Models and Training Setup}
All neural network models share a common input--output structure: they take a batch of 180‑day input windows (each of dimension $T_{in} \times F$, with $F=251$ features) and produce a vector of portfolio allocation weights $w \in \mathbb{R}^N$ ($N=50$) through a final softmax layer.

Training Hyperparameters: All hyperparameters, excluding those listed below are tuned using Optuna; 100 trials per model-loss combination. The following hyperparameters are shared across all neural models:
\begin{itemize}
    \item Optimizer: AdamW.
    \item Batch size: 64 for training.
    \item Gradient clipping: maximum norm = 0.5.
    \item Data shuffling: shuffle=True in the training DataLoader.
\end{itemize}
The two custom loss functions, CustomLossA~\eqref{eq:loss_A} and CustomLossB~\eqref{eq:loss_B}, each contain two regularization weights, $\lambda_{\text{CVaR}}$ and $\lambda_{\text{RP}}$, which are tuned from the set $\{0.001, 0.01, 0.1, 1.0\}$.

\subsection{Transaction Costs}
\label{subsec:transaction_costs}
At each walk-forward step $k$, the portfolio weights that are generated by the model(s) produce a gross daily returns series $r^{\text{gross}, (k)} = R^{(k)}w^{(k)}$, where $R^{(k)}$ denotes the asset returns for $T_{out}$. The cost of rebalancing is incurred on the first day of the output period. Let $w_{\text{prev}}$ be the weights of the previous period ($w_{\text{prev}} = 0$ for $k=1$). The portfolio turnover is, 
\begin{equation}
    \Delta = \vert w^{(k)} - w_{\text{prev}}\vert.
\label{eq:pf_turnover}
\end{equation}
The total cost (as a fraction of the portfolio value) is:
\begin{equation}
    \text{cost}^{(k)} = \frac{1}{2}\sum_{i=1}^N \Delta_{i} \cdot \text{BA-Spread}_i,
\label{eq:ba_cost}
\end{equation}
where $\text{BA-Spread}_{i}$ is the bid-ask spread of asset $i$ on the first day of the current window. The net return on the first day is then,
\begin{equation}
    r_1^{\text{net}, (k)} = (1 + r_1^{\text{gross},(k)}) \cdot (1 - \text{cost}^{(k)}) - 1,
\label{eq:net_rets}
\end{equation}
while the returns for days $2,\dots,60$ remain unchanged (no further cost for holding).

After all 8 steps, the 8 daily return series (each of length 60) are concatenated to form a single 480‑day net return series for the entire validation or test period. All portfolio performance metrics are computed on this concatenated series, ensuring that transaction costs are faithfully accounted for throughout the backtest. The same procedure is applied to all traditional benchmark models for a fair comparison.

\subsection{Evaluation Settings}
\label{subsec:eval_settings}
All final test set evaluations are performed using 30 fixed seeds. No seeds are fixed during hyperparameter tuning; only the test set runs are seeded to enable reproducible statistical comparisons. The seeds are the same for every model, ensuring that paired statistical tests are valid. For each seed and each model-loss combination, we run the full 8‑step walk‑forward procedure described in Section \ref{subsec:wfv}, obtaining a net daily returns series for the entire test period from 2022-01-03 to 2023-11-29 (480 trading days). From this series we compute the following performance metrics:
\begin{itemize}
    \item Compounded return: $R_{\text{total}} = \prod_{t=1}^{480} (1 + r_t) - 1$.
    \item Sharpe ratio: $\text{Sharpe} = (\bar{r} ~/~ \sigma_r) \times \sqrt{252}$.
    \item Sortino ratio: $\text{Sortino} = (\bar{r} ~/~ \sigma_d) \times \sqrt{252}$, with downside deviation $\sigma_d$ computed using a target return of zero.
    \item Maximum Drawdown (MDD): the largest peak‑to‑trough decline over the period.
    \item Calmar ratio: $\text{Calmar} = \frac{(1 + R_{\text{total}})^{\frac{252}{T}} - 1}{|\text{MDD}|}$, where $T=480$.
    \item Omega ratio (threshold $\theta = 0$): $\Omega = \frac{\sum\max(0, r_t)}{\sum \max(0, -r_t)}$.
    \item CVaR ($\alpha = 0.05$): $\operatorname{CVaR}_{\alpha}(r) = \frac{1}{\lfloor \alpha T \rfloor} \sum_{t=1}^{\lfloor \alpha T \rfloor} r_{(t)}$.
\end{itemize}
For the S\&P 500 and equal-weight benchmarks, the same metrics are computed directly from the returns with no transaction costs applied. Sharpe, Sortino, and Calmar ratios are annualized, while compounded return, CVaR, MDD, and Omega are presented as total‑period (non‑annualized) values. Annualization does not change comparative rankings.

\subsection{Hardware}
All computationally intensive tasks (hyperparameter tuning and evaluation) were executed on the Rochester Institute of Technology Research Computing HPC cluster. The cluster nodes are equipped with Intel Xeon processors and NVIDIA A100 GPUs (40GB VRAM). Parallelization across multiple processes was implemented using MPI.

\section{Results}
We evaluated the six candidate models on the out‑of‑sample test period (2022--2023) using 30 random seeds for each model. All reported metrics are computed on daily returns net of transaction costs.

\subsection{Main Performance Comparison} Table \ref{tab:mean_nn} reports the mean (across 30 seeds) for each neural model and a single value for each deterministic benchmark. The AttentionLSTM‑CustomLossB achieved the highest Sharpe ratio (mean $=$ 0.29, 95\% CI [0.21834, 0.36172]), the highest compounded return (+7.86\%), and the best Omega ratio (1.0504). Its MDD (\textminus20.21\%) is the best among all neural models, and is comparable to HRP (\textminus19.80\%) and Equal Risk Contribution with turnover control (\textminus20.06\%). Its CVaR (mean $=-2.86\%$, 95\% CI [\textminus2.942\%, \textminus2.772\%]) is virtually identical to that of the S\&P 500 (\textminus2.84\%). The AttentionLSTM also shows relatively low variability: Sharpe standard deviation of 0.1920 and CVaR standard deviation of 0.2280.

\begin{table*}[htp]
\centering
\setlength{\tabcolsep}{4pt}
\caption{Mean Performance of Neural Network Models Over 2022 \& 2023}
\small
\label{tab:mean_nn}
\begin{tabular}{lccccccc}
\toprule
\textbf{Model/Benchmark} & \textbf{Comp. Return (\%)} & \textbf{Sharpe} & \textbf{Sortino} & \textbf{Omega} & \textbf{Calmar} & \textbf{MDD (\%)} & \textbf{CVaR (\%; $\alpha=5\%$)} \\
\midrule
AttentionLSTM-CustomLossB      & 7.86 & 0.2900 & 0.4599 & 1.0504 & 0.2333 & \textminus20.21 & \textminus2.86 \\
DeformTime-CustomLossB         & 2.61 & 0.1674 & 0.2634 & 1.0329 & 0.0733 & \textminus46.14 & \textminus5.76 \\
TemporalTransformer-CustomLossB & 2.07 & 0.1500 & 0.2492 & 1.0260 & 0.0911 & \textminus23.59 & \textminus3.02 \\
PatchTST-CustomLossB           & \textminus5.96 & \textminus0.0459 & \textminus0.0711 & 0.9928 & \textminus0.1314 & \textminus23.46 & \textminus2.94 \\
VSN-LSTM-CustomLossB          & \textminus8.72 & \textminus0.0755 & \textminus0.1177 & 0.9882 & \textminus0.1472 & \textminus30.02 & \textminus3.29 \\
InvertedAttentionLSTM-CustomLossA & \textminus13.23 & \textminus0.2064 & \textminus0.3304 & 0.9672 & \textminus0.2465 & \textminus27.51 & \textminus3.17 \\
\midrule
\multicolumn{8}{c}{\textit{Benchmarks and Traditional Methods}} \\
\midrule
NestedClusteredOptimization     & \textminus0.64 & 0.0536 & 0.0823 & 1.0089 & \textminus0.0196 & \textminus17.20 & \textminus2.15 \\
S\&P 500                        & \textminus4.52 & \textminus0.0240 & \textminus0.0363 & 0.9960 & \textminus0.0945 & \textminus25.43 & \textminus2.84 \\
HierarchicalRiskParity          & \textminus7.94 & \textminus0.1776 & \textminus0.2831 & 0.9714 & \textminus0.2147 & \textminus19.80 & \textminus2.24 \\
EqualRiskContribution ($\delta=0.2$) & \textminus8.67 & \textminus0.1852 & \textminus0.2961 & 0.9701 & \textminus0.2319 & \textminus20.06 & \textminus2.35 \\
GlobalMinimumVariance           & \textminus9.02 & \textminus0.2573 & \textminus0.3928 & 0.9587 & \textminus0.2874 & \textminus16.84 & \textminus2.10 \\
Equal Weight                    & \textminus9.99 & \textminus0.1999 & \textminus0.3208 & 0.9678 & \textminus0.2531 & \textminus21.23 & \textminus2.55 \\
MeanVariance (Shrinkage)        & \textminus34.21 & \textminus0.6922 & \textminus1.0964 & 0.8901 & \textminus0.4525 & \textminus43.62 & \textminus3.63 \\
\bottomrule
\multicolumn{8}{c}{\footnotesize Note: Only Sharpe, Sortino and Calmar are annualized.}
\end{tabular}
\end{table*}

\subsection{Distribution of Risk‑Adjusted Returns and Tail Risk}
\begin{figure}[ht]
\centering
\includegraphics[width=\columnwidth, height=4in]{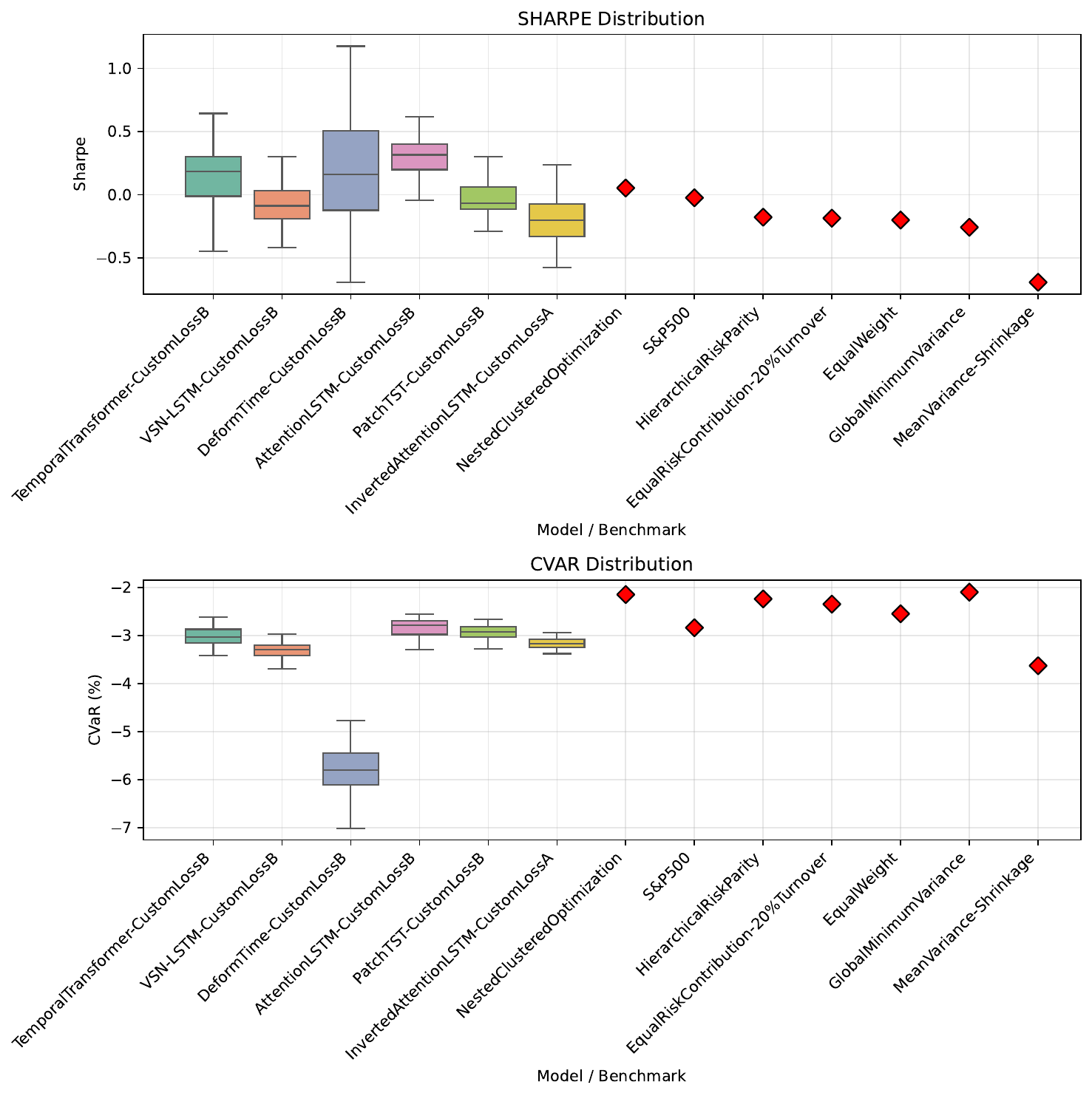}
\caption{Sharpe and CVaR Distribution}
\label{fig:sharpe-cvar}
\end{figure}

Figure \ref{fig:sharpe-cvar} presents box plots of the 30 seed‑wise Sharpe ratios (top) and CVaR values (bottom) for all models and a single value for each deterministic benchmark. The AttentionLSTM clearly stands out: its median Sharpe of 0.3160 is the highest among all models, far exceeding the next best neural network, TemporalTransformer (median $=$ 0.1824), NCO (0.0536), and the S\&P 500 (\textminus0.0240). Moreover, its interquartile range is narrow, indicating consistent performance. Conversely, several neural models (InvertedAttentionLSTM, VSN-LSTM, PatchTST) show median Sharpe ratios below zero, indicating that not every architecture benefits from the custom loss functions.

On the CVaR side, the differences are mostly less dramatic. Most neural models have median CVaR values between $\approx$\textminus3.3\% and $\approx$\textminus2.8\%, similar to the S\&P 500 (\textminus2.84\%). DeformTime, however, exhibits the worst tail risk among all models (median $=$ \textminus5.80\%). The AttentionLSTM's median CVaR (\textminus2.79\%) is comparable to most benchmarks, including Global Minimum Variance (\textminus2.10\%), NCO (\textminus2.15\%), and Equal Risk Contribution (\textminus2.35\%).

\subsection{Trade‑off Between Return and Tail Risk}
\begin{figure}[ht]
\centering
\includegraphics[width=\columnwidth]{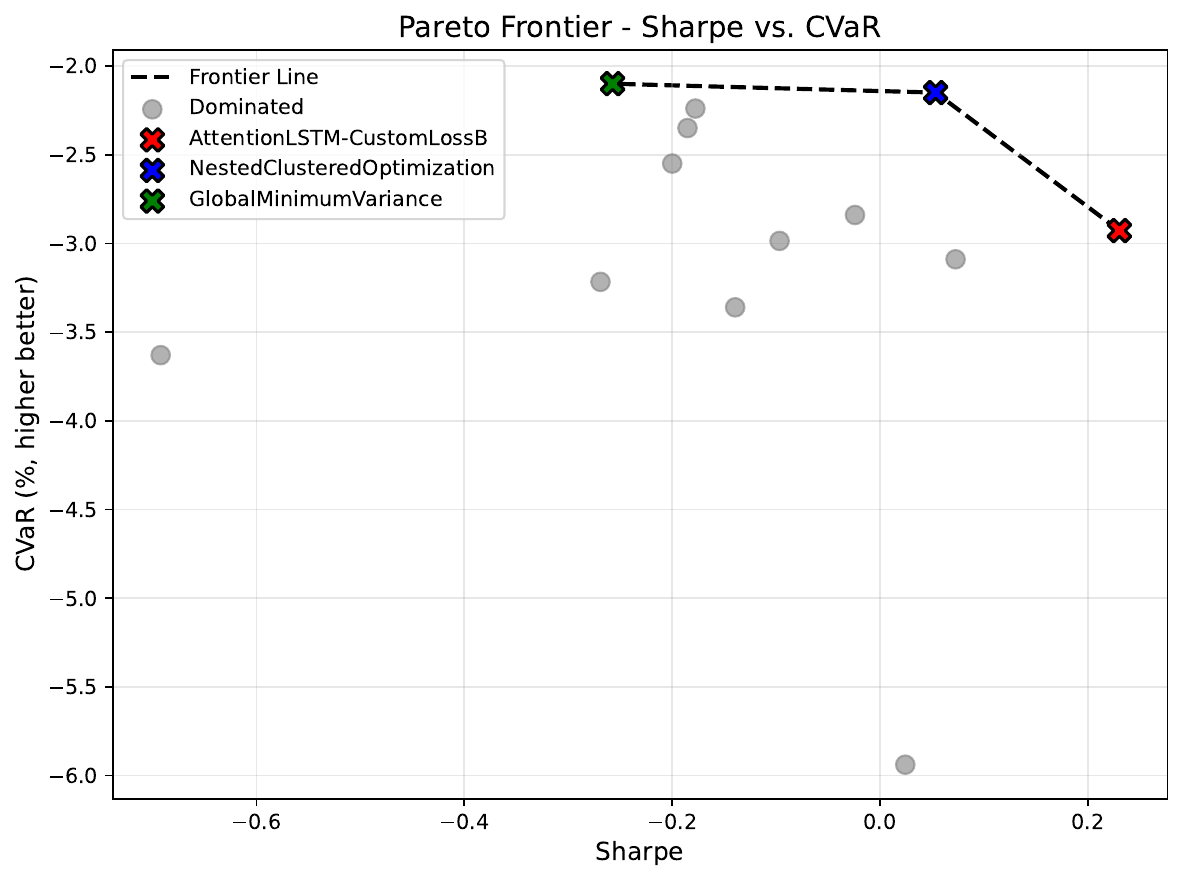}
\caption{Pareto Frontier -- 95\% Lower Confidence Bounds}
\label{fig:test-pareto}
\end{figure}

We constructed a Pareto frontier using the lower bounds of the 95\% confidence intervals of the mean Sharpe and mean CVaR. The frontier (Figure \ref{fig:test-pareto}) shows that the AttentionLSTM lies on the efficient boundary, whereas all other neural models are dominated. In particular, DeformTime and TemporalTransformer offer slightly lower Sharpe with similar or worse tail risk, while VSN‑LSTM and PatchTST are dominated both in Sharpe and CVaR. This confirms that AttentionLSTM-CustomLossB provides the best overall trade‑off between risk‑adjusted return and tail risk.

\subsection{Statistical Significance}
\label{subsec:stat_sig}
We performed one‑sample one‑sided $t$-tests against the S\&P 500 after verifying normality (Shapiro-Wilk test, $p>0.05$ for all models). All reported $p$-values are Bonferroni‑corrected for multiple comparisons. The AttentionLSTM shows a highly significant Sharpe improvement ($t_{29}=8.95$, $p<10^{-8}$). Among other neural models, only TemporalTransformer is significantly better ($p=0.0039$) but with a much smaller effect (mean Sharpe $=$ 0.15); DeformTime is not significant ($p=0.18$); all remaining neural models perform worse ($p=1.0$). For CVaR, the AttentionLSTM was not significantly different from the S\&P 500 ($p > 0.05$), confirming that tail risk did not increase. Against the strongest traditional benchmark, NCO, the AttentionLSTM again yields significantly higher Sharpe ($t_{29}=6.74$, $p=10^{-6}$); no other neural model significantly outperforms NCO (all $p\ge0.26$). Paired $t$-tests confirm that AttentionLSTM significantly outperforms all competing neural models (all $p < 0.05$) except DeformTime ($p=0.53$). 
% Paired $t$-tests confirm that AttentionLSTM significantly outperforms TemporalTransformer ($p=0.013$), VSN‑LSTM, PatchTST, and InvertedAttentionLSTM (all $p<0.0001$), but not DeformTime ($p=0.53$). 
% This demonstrates that its superior performance is consistent across random initializations.

\section{Conclusion}
This research set out to determine whether an end‑to‑end deep learning framework, guided by differentiable financial objectives and a robust evaluation procedure, could produce portfolio allocations that outperform both traditional benchmarks and state‑of‑the‑art neural models.

The test period (2022--2023) was a difficult environment for equities. The S\&P 500 delivered a compounded return of \textminus4.52\% and an annualized Sharpe ratio of \textminus0.02. Despite this, the AttentionLSTM with the Omega-CVaR-RiskParity loss (CustomLossB) achieved a mean annualized Sharpe ratio of 0.29 and a compounded return of +7.86\%. Tail risk (CVaR) increased negligibly relative to the S\&P 500, demonstrating that the model does not simply take on more downside exposure to achieve higher returns. The improvement in Sharpe ratio is statistically robust across 30 random seeds (Bonferroni‑corrected $p<10^{-8}$). 

While the AttentionLSTM significantly outperformed all benchmarks over the full test period, it initially underperformed NCO during the first two quarters of 2022 (the onset of the bear market). The model's performance recovered as the expanding training window incorporated more data from the new regime, leading to superior full-period results. These results show that even a relatively simple recurrent architecture, with our end-to-end framework, can deliver economically meaningful outperformance.

\paragraph{Limitations} This study is primarily a proof of concept for the proposed end-to-end framework, focusing on a single out-of-sample period (2022--2023). While the results demonstrate the framework's potential, they may not generalize to other market samples. The stocks utilized were fixed and consisted of 50 S\&P 500 constituents that were active throughout the study period, which may introduce survivorship bias. Additionally, the framework does not account for short selling, margin requirements, or more sophisticated transaction cost models.

\paragraph{Future Work} Several promising extensions remain. Extending the evaluation to multiple disjoint market periods would provide stronger evidence of generalizability. Incorporating market regime detection (e.g., via Hidden Markov Models) could further improve the framework's adaptability to changing conditions. Expanding the asset universe to include bonds, commodities, and international equities would test the framework's robustness across different asset classes. Finally, extending the framework to long-short strategies would explore its potential beyond the long-only setting.

In summary, embedding financial objectives and portfolio construction directly into both hyperparameter tuning and model training yields economically meaningful and statistically robust performance. These results demonstrate that a well-designed, end-to-end pipeline can serve as a reliable tool for real-world investment. \footnote{The supplementary appendix is available at: \url{https://doi.org/10.5281/zenodo.21925252}. The latest code is available at: \url{https://github.com/rahulkfernandes/Financially-Guided-Deep-Portfolio-Optimization}}

\section{Acknowledgment}
\begin{itemize}
    \item We gratefully acknowledge the use of the RIT Research Computing's HPC cluster at Rochester Institute of Technology, which provided essential computational resources for this research.
    \item This research uses data from the Center for Research in Security Prices (CRSP), accessed through RIT.
\end{itemize}

\bibliographystyle{IEEEtran}
\bibliography{references}

\end{document}